\documentclass[twocolumn,prd,floatfix,preprintnumbers,nofootinbib,superscriptaddress, aps]{revtex4-1}

\usepackage{subcaption}
\usepackage{ragged2e}
\DeclareCaptionJustification{justified}{\justifying}
\usepackage{amssymb,amsmath,verbatim,mathtools,needspace,enumitem,etoolbox,graphicx,physics,microtype,afterpage,bm}
\usepackage{esint}
\usepackage[dvipsnames]{xcolor}
\definecolor{linkcolor}{rgb}{0.0,0.3,0.5}
\usepackage[unicode, colorlinks=true, linkcolor=linkcolor, citecolor=linkcolor, filecolor=linkcolor,urlcolor=linkcolor, pdfusetitle]{hyperref}
\usepackage[all]{hypcap}
\usepackage[T1]{fontenc}
\usepackage[utf8]{inputenc}
\usepackage{tabularx}
\usepackage{cancel}
\usepackage{float}
\newcommand\underrel[3][]{\mathrel{\mathop{#3}\limits_{%
      \ifx c#1\relax\mathclap{#2}\else#2\fi}}}
\usepackage{soul}
\usepackage{lmodern}
\allowdisplaybreaks
\usepackage{tikz}
\usepackage{cleveref}
\usepackage{color}
\usepackage{framed}
\usepackage{hyperref}
\hypersetup{colorlinks, citecolor=bluscuro, linkcolor=black, urlcolor=bluscuro}
\definecolor{rossos}{cmyk}{0,1,1,0.55}
\definecolor{bluscuro}{rgb}{0.15, 0.2, .85}

\newcommand{\be}{\begin{equation}}
\newcommand{\ee}{\end{equation}}

\def\BH{\text{\tiny BH}}

\def\lsim{\mathrel{\rlap{\lower4pt\hbox{\hskip0.5pt$\sim$}}
    \raise1pt\hbox{$<$}}}         
\def\gsim{\mathrel{\rlap{\lower4pt\hbox{\hskip0.5pt$\sim$}}
    \raise1pt\hbox{$>$}}}         

\makeatletter
\newcommand{\subsetsim}{\mathrel{\mathpalette\subset@sim\relax}}
\newcommand{\subset@sim}[2]{%
  \vtop{\offinterlineskip\m@th
    \ialign{\hfil##\cr
     ~$#1\subset$\cr\noalign{\kern0.5pt}\scalebox{0.9}{$#1\sim$}\cr
    }%
  }%
}
\makeatother
\makeatletter
\def\l@subsubsection#1#2{}
\makeatother

\newcommand{\sapienza}{Dipartimento di Fisica, Sapienza Universit\`a di Roma, Piazzale Aldo Moro 5, 00185, Roma, Italy}
\newcommand{\infn}{INFN, Sezione di Roma, Piazzale Aldo Moro 2, 00185, Roma, Italy}

\begin{document}

\title{Dynamical Love numbers of black holes: \\ Theory and gravitational waveforms}

\author{Sumanta Chakraborty}
\email{tpsc@iacs.res.in}
\affiliation{School of Physical Sciences, Indian Association for the Cultivation of Science, Kolkata-700032, India}

\author{Valerio De Luca}
\email{vdeluca@sas.upenn.edu}
\affiliation{Center for Particle Cosmology, Department of Physics and Astronomy,
University of Pennsylvania 209 South 33rd Street, Philadelphia, Pennsylvania 19104, USA}

\author{Leonardo Gualtieri}
\email{leonardo.gualtieri@unipi.it}
\affiliation{Dipartimento di Fisica, Universit\`a di Pisa, 56127 Pisa, Italy}
\affiliation{INFN, Sezione di Pisa, Largo B. Pontecorvo 3, 56127 Pisa, Italy}

\author{Paolo Pani}
\email{paolo.pani@uniroma1.it}
\affiliation{\sapienza}
\affiliation{\infn}


\begin{abstract}
\noindent
In General Relativity, the static tidal Love numbers of black holes vanish identically. Whether this remains true for time-dependent tidal fields---{\it i.e.}, in the case of dynamical tidal Love numbers---is an open question, complicated by subtle issues in the definition and computation of the tidal response at finite frequency.
In this work, we analyze the dynamical tidal perturbations of a Schwarzschild black hole to quadratic order in the tidal frequency. By employing the Teukolsky formalism in advanced null coordinates, which are regular at the horizon, we obtain a particularly clean perturbative scheme. Furthermore, we introduce a response function based on the full solution of the perturbation equation which does not depend on any arbitrary constant.
Our analysis recovers known results for the dissipative response at linear order and the logarithmic running at quadratic order, associated with scale dependence in the effective theory. In addition, we find a finite, nonvanishing conservative correction at second order in frequency, thereby possibly demonstrating a genuine dynamical deformation of the black hole geometry. Although removing any ambiguity in the dynamical tidal response would require a matching with some gauge-invariant coefficient, we assess the impact of these effects on the gravitational-wave phase. These contributions enter at eighth post-Newtonian order, and can be expressed in terms of generic $\mathcal{O}(1)$ coefficients, which have to be matched to the perturbative result.
Regardless of the matching ambiguities, we argue that such corrections are too small to be observable even with future-generation gravitational wave detectors. Moreover, the corresponding phase shifts are degenerate with unknown point-particle contributions entering at the same post-Newtonian order.
\end{abstract}

\preprint{ET-0381A-25}
\maketitle

\section{Introduction}
\label{sec:intro}
\noindent
Since the seminal work of A.E.H.~Love, who studied how Earth's shape deforms under the tidal forces of the Moon and the Sun~\cite{Love1909}, the investigation of an object's gravitational response to an external tidal field has become a fundamental tool for probing its internal properties~\cite{PoissonWill}.
This response is often parametrized in terms of so-called tidal Love numbers~(TLNs).

The tidal response of a black hole~(BH) in General Relativity is quite unique: the \emph{static} TLNs (associated with the response to a static tidal field) vanish identically~\cite{Damour_tidal, Binnington:2009bb, Damour:2009vw, Gurlebeck:2015xpa, Poisson:2014gka, Pani:2015hfa, Landry:2015zfa, LeTiec:2020bos, Chia:2020yla, LeTiec:2020spy,Bhatt:2023zsy}.
This underscores a naturalness problem from the point of view of an effective field theory~\cite{Porto:2016zng, Cardoso:2017cfl} and is a property that does not hold  true for any other object within General Relativity~\cite{Wade:2013hoa,Cardoso:2017cfl, Sennett:2017etc, Mendes:2016vdr, Pani:2015tga, Cardoso:2017cfl, Uchikata:2016qku, Raposo:2018rjn, Cardoso:2019rvt,Berti:2024moe}, for BHs dressed by matter distributions~\cite{Baumann:2018vus,DeLuca:2021ite,DeLuca:2022xlz,Brito:2023pyl,Capuano:2024qhv,Cardoso:2019upw,Cardoso:2021wlq,Katagiri:2023yzm,DeLuca:2024uju,Cannizzaro:2024fpz, DeLuca:2025bph}, for BHs in modified gravity, as well as for asymptotically nonflat geometries~\cite{Cardoso:2017cfl, Cardoso:2018ptl,DeLuca:2024nih, DeLuca:2022tkm, Barbosa:2025uau, Barura:2024uog, nair2024asymptotically-199, Franzin:2024cah}, or in higher spacetime dimensions~\cite{Chakravarti:2018vlt, Chakravarti:2019aup, Pereniguez:2021xcj,Dey:2020lhq, Dey:2020pth, Kol:2011vg,Cardoso:2019vof,Hui:2020xxx,Rodriguez:2023xjd,Charalambous:2023jgq,Charalambous:2024tdj,Charalambous:2024gpf,Ma:2024few}.
Indeed, it can be related to some special symmetries of the perturbations of the Kerr solution in the zero-frequency limit~\cite{Hui:2020xxx, Charalambous:2021kcz, Charalambous:2021mea, Hui:2021vcv, Berens:2022ebl, BenAchour:2022uqo, Charalambous:2022rre, Katagiri:2022vyz, Ivanov:2022qqt, DeLuca:2023mio,
Bonelli:2021uvf,Ivanov:2022hlo,Berens:2022ebl,Bhatt:2023zsy,Sharma:2024hlz,Rai:2024lho}, which has also been investigated beyond the linear regime~\cite{DeLuca:2023mio,Riva:2023rcm,Iteanu:2024dvx} and shown to hold at all orders in perturbation theory~\cite{Kehagias:2024rtz,Combaluzier-Szteinsznaider:2024sgb,Gounis:2024hcm, Lupsasca:2025pnt}.

The leading-order tidal response of a BH in a small-frequency expansion is linear in the frequency of the tidal field and is associated with the tidal heating at the horizon~\cite{Hartle:1973zz,Hughes:2001jr,Chia:2020yla,Chia:2024bwc}, representing a dissipative contribution that is characteristic of BHs.
The next-to-leading-order term is quadratic in the frequency and corresponds to the conservative component of the tidal response. In light of the intriguing result that the static TLNs vanish, determining whether the conservative tidal response of a BH also vanishes at second order is of significant interest.

In this context, dynamical tidal perturbations have been recently studied in~\cite{Bhatt:2024yyz,Katagiri:2022vyz, Ivanov:2022qqt, DeLuca:2023mio,Charalambous:2022rre, Saketh:2023bul,Perry:2023wmm,Chakraborty:2023zed,Ivanov:2024sds,DeLuca:2024ufn,Bhatt:2024yyz,Katagiri:2024wbg,Katagiri:2024fpn,Bhatt:2024rpx}, with different motivations and approaches (see also~\cite{Pitre:2023xsr,HegadeKR:2024agt,Katagiri:2024wbg,HegadeKR:2025qwj} for other recent studies more  focused on the dynamical tidal response of neutron stars). The results of those papers concerning whether the dynamical conservative TLNs of a BH remain zero are inconclusive\footnote{Here we are referring to the dynamical TLNs of a Schwarzschild BH under gravitational perturbations, which is the main focus of this work. For dynamical perturbations of a test scalar field on a Schwarzschild background, the perturbation theory results have been fully matched with the coefficients of the corresponding effective field theory through scattering amplitude techniques, unambiguously showing that the scalar dynamical TLNs of a Schwarzschild BH are nonzero~\cite{Ivanov:2024sds,Caron-Huot:2025tlq}.}, for two main reasons.

First, it is well known that, within BH perturbation theory alone, there are ambiguities in the definition of the TLNs that arise from a coordinate-dependent contamination of the body's response by the subleading terms of the external tidal field (see, e.g.,~\cite{Pani:2015hfa,Pani:2015nua,Gralla:2017djj} for some detailed discussion).
This ambiguity can be resolved using an analytical continuation
of the angular momentum index of the perturbations (or, alternatively, of the number of spacetime dimensions)~\cite{Kol:2011vg,LeTiec:2020bos}. Unfortunately, this is possible only in some simple cases where a general analytical solution for any angular momentum index or any spacetime dimension is available, which is not the case for the dynamical TLNs.  
A more general resolution of the ambiguity involves matching the perturbation to some gauge-invariant coefficient, i.e. a scalar quantity that does not depend on spacetime coordinates. 
For example, one could match with the binding energy of a binary system involving tidally deformed bodies~\cite{Creci:2021rkz} or with the Wilson coefficients of an effective field theory~\cite{Porto:2016zng, Hui:2020xxx}.
Such a matching might require specific coordinate transformations and might be complicated to perform in certain cases. However, once performed, it would unambiguously fix the tidal response.

On top of this known issue, Ref.~\cite{Katagiri:2024wbg} recently pointed out that a further ambiguity affects the dynamical TLNs: within the perturbative scheme of a small-frequency expansion, the dynamical TLNs arising at second order can be contaminated by a free constant entering at linear order (and not affecting the dissipative term). This introduces an ambiguity in the definition of the dynamical TLNs at second order, which requires some arbitrary calibration~\cite{Katagiri:2024wbg, HegadeKR:2024agt}.
While this ambiguity should be resolved by a careful match of the perturbative solution to an effective field theory, it would be desirable to avoid this problem already at the level of BH perturbation theory.

In this work, we present three main results contributing to the ongoing effort toward understanding and quantifying the dynamical tidal response of a BH in General Relativity.

Firstly, we analyze the dynamical tidal response of a Schwarzschild BH up to quadratic order in the tidal-field frequency. At variance with previous studies~\cite{Pitre:2023xsr,HegadeKR:2024agt,Katagiri:2024wbg,HegadeKR:2025qwj}, we consider the Teukolsky formalism (based on perturbation of the  Weyl tensor~\cite{Teukolsky:1973ha, Chia:2020yla, Chakraborty:2023zed}) rather than the metric formalism typically used for static BHs. Most importantly, we consider advanced null coordinates, wherein the BH horizon is regular, at variance with the usual Schwarzschild coordinates used in previous work. While conceptually our approach is equivalent to previous ones, the dynamical equations describing Teukolsky perturbations in advanced null coordinates are different from the usual Regge-Wheeler perturbations. This allows us to recast the perturbations order by order in the frequency in a particularly convenient form. 
Moreover, we define the tidal response function in terms of the full solution of the perturbative equation, including both the homogeneous and the particular solutions. We shall show that, after imposing regularity boundary conditions, this response function is independent of arbitrary constants associated to the homogeneous solution.
Our approach correctly reproduces
known results for the dissipative response at linear order in the frequency and for the logarithmic
correction at quadratic order, related to the running of the TLNs within the perturbative
approach~\cite{Saketh:2023bul}.

Secondly, we perform a  post-Newtonian~(PN) computation of the gravitational-wave~(GW) signal emitted by a BH binary including dynamical tidal effects at the leading PN order, which introduces an 8PN correction to the GW phase~\cite{Pitre:2023xsr}.

Finally, although in this paper we do not perform the exact matching of the TLNs computed within BH perturbation theory to the waveform coefficients, we estimate whether such coefficients can be constrained by GW observations. 
As it turns out, regardless of the matching ambiguity,  their effect on the GW phase is too small to be detected even for the loudest events expected in future detectors such as LISA~\cite{LISA:2024hlh} or the Einstein Telescope~(ET)~\cite{Abac:2025saz, Branchesi:2023mws} (unless the actual  matching procedure leads to unnaturally large numerical factors, which would be surprising from an effective field theory perspective).
Therefore, besides their theoretical interest and the subtleties in a proper matching into the waveform, the dynamical TLNs of BHs should not have any observational impact in the foreseeable future. 
The situation may be qualitatively different for neutron stars, where dynamical tidal effects could be enhanced by additional compactness-dependent contributions, a possibility we leave for future investigation.

Throughout this work we set geometrical units \( G = c = 1 \), and adopt the mostly-plus metric signature. In Cartesian coordinates, the flat Minkowski metric is given by $\eta_{\mu \nu} = \mathrm{diag}(-1, 1, 1, 1)$.
Greek letters \( \mu, \nu, \alpha, \ldots \) denote spacetime indices, while Latin letters \( i, j, k, \ldots \) are reserved for spatial components only.

\section{Dynamical TLNs from a small-frequency expansion}
\label{sec:GR}
\noindent
In this Section we will use BH perturbation theory to compute the conservative and dissipative dynamical response of a Schwarzschild BH. We will often refer to the former simply as dynamical TLNs, and to the latter as tidal dissipation numbers.

The computation of TLNs and tidal dissipation numbers follows from the determination of the tidal response function. For this purpose, one solves the relevant perturbation equations and determines the associated perturbation variables. In the case of a static and spherically symmetric spacetime, generically these perturbation variables are considered to be the perturbations of the metric components~\cite{Binnington:2009bb,Damour:2009vw}, $g_{tt}$ and $g_{t\phi}$ for the even-parity and odd-parity TLNs, respectively, while for spinning BHs the perturbation variables are naturally the Weyl scalars arising from the Newman-Penrose formalism~\cite{Chia:2020yla}. 

Irrespective of the origin of the perturbation variable, after solving the relevant equation, the tidal response function can be extracted from the asymptotic behavior of the perturbation. 
For static perturbations, this is indeed an asymptotic expansion, while for dynamical perturbations with frequency $\omega$, the expansion is performed in such a way that any relevant length scales $R$ are smaller compared to $\omega^{-1}$. In either of these situations, the asymptotic expansion typically has a growing part and a decaying part. Schematically, for a suitable perturbation variable $\Psi$,
\begin{align}\label{gen_asymp}
\Psi&\approx G(\ell,m_{z},s,\omega,h_{i})\left(\frac{r}{R}\right)^{\ell-s} \nonumber \\
& +D(\ell,m_{z},s,\omega,h_{i})\left(\frac{r}{R}\right)^{-\ell-s-1}\,,
\end{align}
where $s$ is the spin of the perturbation, $\ell$ and $m_{z}$ denote the angular and azimuthal numbers (related to the eigenvalues of the angular harmonics), respectively, and $h_{i}=\{m,J,\cdots\}$ are the `hairs', {\it i.e.} the quantities characterizing the (stationary) compact object, with $m$ being the mass and $J$ the angular momentum. Here, the growing mode ($\sim r^{\ell-s}$) is associated with the strength of the external tidal field $G$, while the decaying mode ($\sim r^{-\ell-s-1}$) corresponds to the response $D$ of the compact object to the tidal field, both normalized with respect to the characteristic scale of the problem at hand, $R$. Therefore, one schematically defines the response function as\footnote{Let us stress that, when switching on the frequency dependence in the external perturbations, one would also induce tail effects in the full solution, associated to the backscattering of waves on the BH background and sourced by the branch cut at $\omega = 0$, and running along the negative frequency imaginary axis, in the retarded Green function~\cite{Price:1971fb}. This contribution could potentially overlap with the dynamical tidal response~\cite{DeLuca:2024ufn}, and isolating them would again require a matching to a worldline effective theory.}
\begin{align}\label{gen_resp}
\,_{s}\mathcal{F}_{\ell m_z}=\frac{D(\ell,m_{z},s,\omega,h_{i})}{G(\ell,m_{z},s,\omega,h_{i})}\,.
\end{align}
In general, the above response function is complex, with its real part providing the conservative piece, defined as the TLNs $\,_{s}k_{\ell m_{z}}$, while the imaginary part is the dissipative component, defined as the dissipation numbers $\,_{s}\nu_{\ell m_{z}}$. Namely, 
\begin{align}\label{gen_TLN_diss}
\,_{s}k_{\ell m_{z}}\equiv\frac{1}{2}\textrm{Re}\left(\,_{s}\mathcal{F}_{\ell m_{z}}\right)\,;\quad \,
\,_{s}\nu_{\ell m_{z}}\equiv \textrm{Im}\left(\,_{s}\mathcal{F}_{\ell m_{z}}\right)\,.
\end{align}
Note that, to define the response function, we shall use the entire solution of the perturbation equation, including both homogeneous and inhomogeneous parts.

In our approach, we will first determine the relevant perturbation equations, and then we will solve them with appropriate boundary conditions (which in the case of BHs demands regularity at their horizon). After that, we will take the asymptotic expansion of the full solution and hence determine the functions $G(\ell,m_{z},s,\omega,h_{i})$ and $D(\ell,m_{z},s,\omega,h_{i})$, leading to the response function and eventually to the TLNs and dissipation numbers through Eq.~\eqref{gen_TLN_diss}.

Our aim, in this Section, is to compute the dynamical tidal response  of a Schwarzschild BH order by order in a $m\omega\ll1$ expansion. In particular, we are interested in the leading-order conservative term at ${\cal O}(m^2\omega^2)$, which has been the focus of recent studies~\cite{Perry:2023wmm,Pitre:2023xsr,HegadeKR:2024agt,Katagiri:2024wbg,HegadeKR:2025qwj}. Unlike previous attempts, we start with the Teukolsky formalism and write down the following decomposition of the Weyl scalar (for the time being, we keep the rotation of the BH to be nonzero and assume arbitrary frequency dependence): 
\begin{widetext}
\begin{equation}\label{Weyl_Decomp}
\rho^{-s+|s|}\Psi_{s-|s|}(v,r,\theta,\bar{\phi})=\int d\omega \sum_{\ell, m_{z}}e^{-i\omega v}e^{-im_{z}\bar{\phi}}\,\,_{s}R_{\ell m_{z}}(r)\,_{s}S_{\ell m_{z}}(\theta,\bar{\phi})\,.
\end{equation}
\end{widetext}
Here, $\rho=r-ia\cos \theta$, $a=J/m$ is the BH effective spin parameter and, for $s=\pm 2$, $\Psi_{0}$ and $\Psi_{4}$ are perturbed Weyl scalars, capturing the perturbations about the gravitational Kerr background. 
Importantly, we will consider the advanced null coordinates $(v,r,\theta,\bar\phi)$, which are regular at the horizon. The coordinates $v$ and $\bar\phi$ are related to the Boyer-Lindquist time and azimuthal coordinates $t$ and $\phi$ through the following relations:
\begin{align}
    dv=dt+\frac{(r^{2}+a^{2})}{\Delta}dr\,; \qquad
    d\bar{\phi}=d\phi+\frac{a}{\Delta}dr\,,
\end{align}
with $\Delta(r)=r^{2}-2mr+a^{2}$. The location of the (outer and inner) horizon corresponds to the roots of $\Delta$, namely $r_\pm=m\pm\sqrt{m^2-a^2}$, as in Boyer-Lindquist coordinates.  The above decomposition of the Weyl scalars leads to separability of the gravitational perturbation equations, with the angular part given by spin-weighted spheroidal harmonics, $\,_{s}S_{\ell m_{z}}$, and the radial part $\,\,_{s}R_{\ell m_{z}}$ satisfying the Teukolsky equation. For $s=-2$, in these coordinates, such equation reads\footnote{Using the $\Upsilon$ variable defined in~\cite{Teukolsky:1973ha} and accounting for a typo in the original Teukolsky's formula as pointed out in \cite{Chatziioannou:2012gq,Chia:2020yla}.}~\cite{Teukolsky:1973ha,Chia:2020yla,Chakraborty:2023zed} (in the following we drop the subscript $s$ from the radial solution $\,_{s}R_{\ell m_z}$; its behavior for generic $s$ is shown in Appendix~\ref{app:genericspin})
\begin{widetext}
\begin{align}\label{gen_grav_rad}
&\dfrac{d^{2} R_{\ell m_{z}}}{dz^{2}}+\left[\frac{2iP_{+}-1}{z}-\frac{2iP_{-}+1}{1+z}-2i\omega\left(r_{+}-r_{-}\right)\right]\dfrac{d R_{\ell m_{z}}}{dz}
+\left[\frac{4iP_{-}}{(1+z)^{2}}-\frac{4iP_{+}}{z^{2}}-\frac{A_{+}+iB_{+}}{z}+\frac{A_{-}+iB_{-}}{(1+z)}\right] R_{\ell m_{z}}=0\,,
\end{align}
\end{widetext}
where we have introduced the rescaled radial coordinate $z=(r-r_{+})/(r_{+}-r_{-})$, along with the following definitions,
\begin{align}
\label{PBA}
P_{\pm}&=\frac{am_{z}-2mr_{\pm}\omega}{r_{+}-r_{-}}\,;
\quad B_{\pm}=2\omega r_{\pm}\,, \nonumber
\\
A_{\pm}&=E_{\ell m_{z}}-2-2m_{z}a\omega+a^{2}\omega^{2}\,.
\end{align}
Here, $E_{\ell m_{z}} = \ell(\ell+1) - s(s+1)$ are the eigenvalues of the angular equation satisfied by the spin-weighted spheroidal harmonics, which depend on the spin of the perturbation. Considering the differential equation for the radial function, one expects to solve it either in an exact manner or upon assuming certain approximations. The resulting solution must be subjected to appropriate boundary conditions at the surface of the compact object, e.g., purely ingoing at the BH event horizon. After imposing the relevant boundary conditions, one expands the radial perturbation function asymptotically, obtaining the following behavior: 
\begin{align}
\label{defResponse}
R_{\ell m_{z}} (r) \propto \left(\frac{r}{r_+}\right)^{\ell+2}\left[\underbrace{1+\cdots}_{\rm tidal\,part}+\underbrace{_{-2}\mathcal{F}_{\ell m_{z}}\left(\frac{r_+}{r}\right)^{2\ell+1}+\cdots}_{\rm response}\right]\,,
\end{align}
where $_{-2}\mathcal{F}_{\ell m_{z}}$ is the response function for $s = -2$ perturbations.

One may easily notice that the tidal field has several sub-leading pieces, denoted by `$\cdots$' in the above expression. It is possible that one of the sub-leading pieces of the tidal field also scales as $r^{-2\ell-1}$, identically to the response, leading to possible ambiguities in the extraction of the response coefficient. These ambiguities can be resolved, either by matching the response function calculated above with the coefficients of an effective field theory action, or by performing analytic continuation, {\it i.e.} promoting the angular multipole to be complex, $\ell\in \mathbb{C}$, and taking $\ell\to \mathbb{Z}^{+}$ at the very end. 
Furthermore, while the extraction of the response function can involve coordinate (gauge) ambiguities---especially when using gauge-dependent quantities---this issue is minimized here since we work with Weyl scalars, which are intrinsically gauge invariant. Nevertheless, some coordinate dependence remains due to the choice of radial coordinate in the asymptotic expansion.

In the following, we will specialize to the Schwarzschild case. Substituting $a=0$ in the above expressions, we obtain
\begin{align}
P_{+}=-2m\omega\,;
\,\,\,
B_{+}= -2P_{+}\,;
\,\,\,
A_{\pm}=(\ell-1)(\ell+2)\,,
\end{align}
while $P_{-}=B_{-} = 0$. As evident, the radial equation, in the Schwarzschild limit, becomes independent of the azimuthal number $m_{z}$ and hence we will omit it in the subsequent computations. Therefore, the radial differential equation on a Schwarzschild background boils down to the form 
\begin{align}
\label{EOMs-2l2}
&z(1+z)\dfrac{d^{2}R_{\ell}}{dz^{2}}-\left[(1+2z)+4im\omega(1+z)^{2}\right]\dfrac{dR_{\ell}}{dz}
\nonumber
\\
&-\Big[(\ell-1)(\ell+2)+4im\omega\frac{(1+z)(z-2)}{z}\Big]R_{\ell}=0\,.
\end{align}
Importantly, there is no term of $\mathcal{O}(m^{2}\omega^{2})$ in the above expression, which is unlike any other approaches in the context of BH perturbation theory, since terms of $\mathcal{O}(m^{2}\omega^{2})$ are present in the standard metric approach. Indeed, this feature is  specific to the perturbations of a Schwarzschild BH expressed in terms of Weyl scalars in advanced null coordinates. Teukolsky equations, expressed either in the Boyer-Lindquist or in the retarded null coordinates, have terms of $\mathcal{O}(m^{2}\omega^{2})$~\cite{Teukolsky:1974yv}. 

The absence of ${\cal O}(m^{2}\omega^{2})$ terms in the equation makes the perturbative approach particularly clean. Indeed, expanding the radial function as
\begin{equation}
R_{\ell}=R^{(0)}_{\ell}+m\omega\,R^{(1)}_{\ell}+m^{2}\omega^{2}\,R^{(2)}_{\ell}+\cdots\,,
\end{equation}
allows us to obtain the following recurrence relations among the radial functions appearing order by order in $m\omega\ll1$:
\begin{align}
\mathcal{D}_{0}R^{(0)}_{\ell}&=0\,,
\label{basic}
\\
\mathcal{D}_{0}R^{(1)}_{\ell}&=4i\mathcal{D}_{1}R^{(0)}_{\ell}\,,
\label{recursion}
\\
\mathcal{D}_{0}R^{(2)}_{\ell}&=4i\mathcal{D}_{1}R^{(1)}_{\ell}\,,
\label{recursion2}
\end{align}
where
\begin{align}
\mathcal{D}_{0}&=z(1+z)\dfrac{d^{2}}{dz^{2}}-(1+2z)\dfrac{d}{dz}-(\ell-1)(\ell+2)\,, 
\\
\mathcal{D}_{1}&=(1+z)^{2}\dfrac{d}{dz}-\left(\frac{2}{z}+1-z\right)\,.
\end{align}
Remarkably, this series continues, so that, at any order $(m\omega)^{j}$, we get the inhomogeneous equation\footnote{Although we will truncate the expansion at ${\cal O}(m^2\omega^2)$, the recurrence relation in Eq.~\eqref{gen_rec} suggests that it might be possible to compute corrections at any order, and then possibly re-sum the $m\omega\ll1$ series.} 
\begin{align}
\mathcal{D}_{0}R^{(j)}_{\ell}&=4i\mathcal{D}_{1}R^{(j-1)}_{\ell}\,.
\label{gen_rec}
\end{align}
Thus, the low-frequency expansion of the Teukolsky radial function for Schwarzschild BH in advanced null coordinates is determined by only \emph{two} differential operators, $\mathcal{D}_{0}$ and $\mathcal{D}_{1}$. 
Note also that the $i$ factor in front of the source term in the recurrence relation in Eq.~\eqref{gen_rec} shows that even-order corrections are real (hence contributing to the conservative piece of the tidal response) while odd-order corrections are purely imaginary (hence contributing to the dissipative part of the tidal response).

At zeroth order in the frequency, Eq.~\eqref{basic} can be solved in closed form in terms of the associated  Legendre polynomial, 
\begin{align}\label{sol_zero}
R^{(0)}_{\ell}=z(1+z)\Big[c_{1}P_{\ell}^{2}(1+2z)+c_{2}Q_{\ell}^{2}(1+2z)\Big]\,,
\end{align}
where $\delta g_{tt}=-f R_\ell/z(1+z)$, with $f(r)=1-(2m/r)$, is the perturbation of the $g_{tt}$ component of the Schwarzschild background. The associated Legendre polynomials can be expressed in terms of hypergeometric functions, with arguments $1/(1+z)$ or $2m/r$. Therefore, the radial Teukolsky function reads 
\begin{widetext}
\begin{align}\label{tiderespzero}
R^{(0)}_{\ell}=f^2\left(\frac{r}{2m}\right)^{2}\Bigg[\mathcal{A}\,\underbrace{\left(\frac{r}{2m}\right)^{\ell}\,_{2}F_{1}\left(-\ell+2,-\ell,-2\ell;\frac{2m}{r}\right)}_{\rm tidal \,part}+\mathcal{B}\,\underbrace{\left(\frac{r}{2m}\right)^{-\ell-1}\,_{2}F_{1}\left(\ell+1,\ell+3,2+2\ell;\frac{2m}{r}\right)}_{\rm response}\Bigg]\,,
\end{align}
\end{widetext}
where the growing mode ($\sim r^{\ell}$) corresponds to the tidal field, while the decaying mode ($\sim r^{-\ell-1}$) is associated to the response function. 
The arbitrary constants $\mathcal{A}$ and $\mathcal{B}$ are related to the arbitrary constants $c_{1}$ and $c_{2}$ in Eq.~\eqref{sol_zero} as
\begin{align}
\mathcal{A}&=c_{1}\frac{\Gamma(1+2\ell)}{\Gamma(\ell+1)\Gamma(\ell-1)}\,,
\label{relation1}
\\
\mathcal{B}&=\frac{\Gamma(\ell+3)\Gamma(\ell+1)}{2\Gamma(2+2\ell)}\left\{c_{2}+c_{1}\frac{\tan(\pi \ell)}{\pi}\right\}\,.
\label{relation2}
\end{align}
Note that, for integer values of $\ell$, the arbitrary constant $\mathcal{B}$ can be simply related to $c_{2}$. Given this solution for the zeroth-order radial function $R^{(0)}_{\ell}$, one should substitute the same in Eq.~\eqref{recursion} and hence obtain the first-order radial function $R^{(1)}_{\ell}$, and so on at any given order. Unfortunately, obtaining an analytical solution for the first-order radial function $R^{(1)}_{\ell}$ for generic $\ell$ does not seem to be possible; hence, we will specialize to the most interesting $\ell=2$ case (the procedure described below can be straightforwardly performed for any integer $\ell$).

Before performing such computation, let us however briefly discuss the near-horizon limit of the zeroth-order radial perturbation equation, presented in Eq.~\eqref{sol_zero}. For this purpose, we first express it as follows
\begin{align}
R^{(0)}_{\ell}&=c_{1}\frac{\Gamma(\ell+3)}{2\Gamma(\ell-1)}z^{2}(1+z)^{2}\,_{2}F_{1}(\ell+3,2-\ell,3;-z)
\nonumber
\\
&+c_{2}\frac{\Gamma(\ell+3)\Gamma(\ell+1)}{2}(1+z)^{2}z^{-\ell-1}
\nonumber
\\
&\qquad \times\,_{2}F_{1}\left(\ell+1,\ell+3,2+2\ell;-\frac{1}{z}\right)\,.
\end{align}
Imposing regularity at the BH horizon sets $c_{2}=0$; hence, the zeroth-order radial function reduces to 
\begin{align}\label{sol_zeroBH}
R^{(0)}_{\ell}=z(1+z)c_{1}P_{\ell}^{2}(1+2z)\,,
\end{align}
which takes a simple polynomial form for integers $\ell$. From such solution one can compute the asymptotic behavior, yielding $\,_{-2}R^{(0)}_{2}\sim z^{4}$. Thus, in the absence of any decaying term, it follows that the response function itself vanishes, see Eqs.~\eqref{gen_asymp} and~\eqref{gen_resp}. Hence, from Eq.~\eqref{gen_TLN_diss} it follows that, for a Schwarzschild BH,
\begin{equation}
\,_{-2}k^{(0)}_{2}=0=_{-2}\nu_{2}^{(0)}\,,
\end{equation}
where the superscript refers to the order in the $m\omega\ll1$ expansion (zeroth-order in this case).
This corresponds to the well-known result that Schwarzschild BHs have vanishing static TLNs~\cite{Damour_tidal,Binnington:2009bb,Damour:2009vw}. Notice that the static dissipation numbers also vanish identically for a Schwarzschild BH, as expected on general grounds, since dissipative terms should be frequency dependent.  

For the linear-in-frequency behavior, one can substitute the above zeroth-order solution back to Eq.~\eqref{recursion} and hence obtain the following solution for the first-order radial perturbation
\begin{align}\label{sol_oneBH}
&R^{(1)}_{2}=-12 i c_{3}z^{2}(1+z)^{2}+\frac{i c_{4}}{2}(1+2z)[6z(1+z)-1]
\nonumber
\\
&-6 i c_{4} z^{2}(1+z)^{2}\ln \left(\frac{1+z}{z}\right)
\nonumber
\\
&-ic_{1}\{24z^{2}(1+z)^{2}\ln(1+z)+2z^{3}[28+z(47+20z)]\}\,,
\end{align}
where the third line of Eq.~\eqref{sol_oneBH} represents the particular solution, proportional to the  amplitude of the static tidal field $c_1\in \mathbb{R}$, while the first two lines describe the homogeneous solutions, with integration constants $c_3$ and $c_4$. As previously noted, the presence of the $i$ factor in front of $c_1$---originating from the $4i$ term on the right-hand side of Eq.~\eqref{recursion}---leads to a purely imaginary contribution in the particular solution, as expected.
Furthermore, due to the presence of logarithmic terms, regularity at the horizon\footnote{Due to the $z^{2}$ factor in front of the logarithm term, $R^{(1)}_2$ and its first derivative are well behaved, but the second derivative is not. Furthermore, $(R^{(1)}_2/z)$ is related to the metric perturbation, thus regularity of metric perturbation will also require $c_{4}=0$. Finally, the total perturbation will have terms of the form $m\omega \ln z$, which will arise from the outgoing mode at the horizon, and hence must vanish.} demands $c_{4}=0$, thus leaving only two arbitrary constants, given by $c_{1}$ and $c_{3}$. It turns out that, to linear order in $m\omega$, the total solution that is regular at the horizon has the following asymptotic behavior
\begin{align}\label{asymplinear}
&R^{(0)}_{2}+m\omega R^{(1)}_{2}\approx-40c_{1}(im\omega)\left(\frac{r}{2m}\right)^{5}-\frac{1}{3}\Big[36c_{1}
\nonumber
\\
&+282i c_{1}m\omega +36i c_{3}m\omega +72ic_{1}m\omega \ln \left(\frac{r}{2m}\right)\Big]\left(\frac{r}{2m}\right)^{4}
\nonumber
\\
&+\dots-\frac{8im\omega c_{1}}{10}\left(\frac{2m}{r}\right)+\mathcal{O}(r^{-2})\,,
\end{align}
where the dots indicate terms scaling as $(r/2m)^{n}$, with $n=3,2,1,0$, which are irrelevant for the extraction of the tidal response. Therefore, the response function up to linear-in-frequency terms can be read using the definition of Eq.~\eqref{defResponse} and becomes
\begin{align}\label{response}
\,_{-2}{\mathcal F}_{2}(\omega)=\frac{im\omega}{15}+\mathcal{O}(m^{2}\omega^{2})\,.
\end{align}
Since the above response function is purely imaginary, it follows that the linear-in-frequency TLN for the $\ell=2$ mode vanishes identically, while giving nonzero dissipation number, $\,_{-2}\nu_{2}^{(1)}=1/15$. It is straightforward to generalize this result to generic $\ell$ modes, yielding~\cite{Chia:2020yla, Charalambous:2021mea} 
\begin{align}
\,_{-2}k_{\ell}&=0+\mathcal{O}(m^{2}\omega^{2})\,,
\\
\,_{-2}\nu_{\ell}&=\frac{2(\ell+2)!(\ell-2)!(\ell!)^{2}}{(2\ell+1)!(2\ell)!}+\mathcal{O}(m^{2}\omega^{2})\,.
\label{dissi}
\end{align}
At this point, it is possible to use the first-order solution shown in Eq.~\eqref{sol_oneBH} as a source term in Eq.~\eqref{recursion2}, and hence solve for the second-order radial perturbation, which reads
\begin{widetext}
\begin{align}
R^{(2)}_{2} & = -12 c_5 z^2 (1+z)^2 + \frac{1}{2} c_6 \left[12 z^3+18 z^2 - 12 (1+z)^2 z^2 \ln \left(\frac{1+z}{z}\right)+4 z-1\right] \nonumber \\
& +2  c_3 z^2 \left[z \left(20 z^2+47 z+28\right)+12 (1+z)^2 \ln (1+z)\right] + 48 c_1 z^2 (1+z)^2 \text{Li}_2(1+z) \nonumber \\
& +\frac{1}{35} c_1 \left[4 (1+z)^2 \left(700 z^3+459 z^2+420 z^2 \ln \left( 1+z\right)+210 z-35\right) \ln (1+z) \right. \nonumber \\
& \left. +z \left(2400 z^5+7060 z^4+4631 z^3-996 z^2-630 z+140\right)\right]\,,
\end{align}
\end{widetext}
in terms of the dilogarithm function $\text{Li}_2 (1+z)$. 
It is clear from our previous analysis that there is a particular solution, depending on $(c_{1},c_{3})$, along with a linear combination of the two homogeneous solutions, involving two further constants, say $c_5$ and $c_6$. The homogeneous solution involving a logarithmic term should be set to zero by regularity at the horizon, enforcing $c_6 = 0$. The subsequent asymptotic expansion of the radial perturbation, up to second-order in frequency, yields
\begin{widetext}
\begin{align}
&R_2\approx \frac{480}{7} c_1 m^2 \omega^2 \left(\frac{r}{2m}\right)^6+\frac{4}{7} \left(\frac{r}{2m}\right)^5 \left[140 c_1 m^2 \omega^2 \ln \left( \frac{r}{2m} \right)+353 c_1 m^2 \omega^2+70  c_3 m^2 \omega^2-70 i c_1 m \omega\right] \nonumber \\
&+\frac{1}{35} \left(\frac{r}{2m}\right)^4 \left[840 c_1 m^2 \omega^2 \ln^2\left( \frac{r}{2m} \right)+7436 c_1 m^2 \omega^2 \ln \left( \frac{r}{2m} \right)+840 c_3 m^2 \omega^2 \ln \left( \frac{r}{2m} \right) - 280 \pi ^2 c_1 m^2 \omega^2\right. \nonumber \\
& \left.   +7431 c_1 m^2 \omega^2+3290 c_3 m^2 \omega^2-420 c_5 m^2 \omega^2-840 i c_1 m \omega \ln \left( \frac{r}{2m} \right)-3290 i c_1 m \omega-420 i c_3 m \omega-420 c_1\right] \nonumber \\
& + \dots  -\frac{2  \left(73 c_1 m^2 \omega^2-14  c_3 m^2 \omega^2+14  ic_1 m \omega \right)}{35} \left(\frac{2m}{r}\right)+\mathcal{O}(r^{-2})\,,
\label{R_second_order}
\end{align}
\end{widetext}
from which it is possible to extract the following response function 
\begin{align}
\label{response-dyn}
\,_{-2}\mathcal{F}_{2}(\omega)=\frac{im\omega}{15}+\frac{2m^{2}\omega^{2}}{15}\left[\ln \left(\frac{r}{2m}\right)+\frac{137}{21}\right]\,,
\end{align}
as a ratio of the $r^{-1}$ term with the $r^{4}$ term, and then expanding in powers of $m\omega$. 

Some important comments are in order. Firstly,  this result does not depend on the arbitrary constants arising at lower order, at variance with previous approaches. Indeed, while the $c_3$ and $c_5$ constants of the homogeneous solutions at lower order appear in the full solution~\eqref{R_second_order}, they do not affect our response function~\eqref{response-dyn} to quadratic order in the frequency.
We attribute this feature to 
our definition of the response function in terms of the entire solution (at variance with other approaches~\cite{Pitre:2023xsr,Katagiri:2024wbg, HegadeKR:2024agt}), to the condition of regularity at the BH horizon, which sets the coefficients ($c_2,c_4$) to zero, as well as to the use of the advanced null coordinates, which are particularly convenient to solve the iterative problem in frequency.

Secondly, one may easily notice the presence of a logarithmic term in the tidal response, highlighting a running behavior. Such running is associated to the existence of UV divergences in the corresponding graviton scattering processes (see, e.g., Ref.~\cite{Ivanov:2024sds}), which are expected to arise at this order in the tidal response. Indeed, a comparison with Ref.~\cite{Saketh:2023bul}, where dynamical TLNs have been extracted through a matching between scattering amplitudes and BH perturbation theory, confirms the running behavior, with the corresponding beta function identical to the dissipative coefficient $_{-2}\nu_{2}^{(1)}$ at the lower order in frequency. Such nontrivial dependence seems to be valid for generic $\ell$'s and is expected to appear through a renormalization procedure in the angular multipole moment $\ell$, in analogy with the results of the Mano–Suzuki–Takasugi (MST) approach~\cite{Mano:1996vt, Mano:1996gn, Mano:1996mf, Sasaki:2003xr}. A deeper investigation of this feature is left to future work\footnote{Let us also mention that, contrarily to the case of BHs where logarithmic terms appear at order $\mathcal{O}(m^2 \omega^2)$, for the case of perfect-fluid neutron stars a running behavior is expected to appear only at higher orders in frequency, in agreement with Refs.~\cite{Pitre:2023xsr,Pitre:2025qdf}. This feature is due to the absence of a tidal dissipation term in the tidal response of perfect-fluid stars at linear order in frequency. At higher order, it is expected that tail effects produce logarithmic runnings also in the perfect-fluid neutron star case.}.

Finally, Eq.~\eqref{response-dyn} suggests that Schwarzschild BHs might have nonzero TLNs at quadratic order in the frequency, while the dissipative term does not have a $m^{2}\omega^{2}$ piece and hence is given by Eq.~\eqref{dissi} up to that order in frequency. The nonzero TLN for the $\ell=2$ mode reads 
\begin{equation}
\,_{-2}k_{2}=\frac{m^{2}\omega^{2}}{15}\left[\ln \left(\frac{r}{2m}\right)+\frac{137}{21}\right]\,.
\label{eq:DTLN}
\end{equation}
The discussion above focused on gravitational perturbations, specifically for spin-weight $s=-2$. In Appendix~\ref{app:genericspin}, we extend the analysis to generic spin-$s$ perturbations of a Schwarzschild BH, finding qualitatively similar behavior. We emphasize that the computation can be straightforwardly generalized also to induced moments with higher values of $\ell$ and different parity.

Before concluding this Section, it is important to stress that the result of Eq.~\eqref{eq:DTLN} may suffer from ambiguities associated to coordinate invariance. In particular, one could perform a coordinate transformation to potentially affect the tidal response of the BH at large distances~\cite{Pani:2015hfa,Pani:2015nua,Gralla:2017djj}.
In order to obtain a truly gauge- and coordinate-independent characterization of the tidal response, one must match the General Relativity computation to an observable quantity in a well-defined gauge-invariant framework, such as the worldline effective field theory  approach~\cite{Goldberger:2005cd, Goldberger:2020fot,Porto:2016zng,Hui:2020xxx}. In this context, the TLNs are identified with Wilson coefficients of higher-derivative operators in the effective action. These coefficients are physical and gauge invariant by construction, and they govern how the tidal response enters into observables, such as the gravitational waveform. As we will show later, these coefficients directly affect the emitted gravitational radiation through their imprint on the waveform phase and amplitude.
A full matching between the General Relativity computation and the worldline theory would require evaluating appropriate graviton scattering amplitudes in the BH background to extract the corresponding Wilson coefficients, which we leave to future work.

This implies that Eq.~\eqref{eq:DTLN} cannot be trusted as the definite prediction for the dynamical TLNs of Schwarzschild BHs entering observable quantities (such as GW signals, see next Section). 
The actual coefficient appearing in the waveform might be related to $\,_{-2}k_{2}$ in Eq.~\eqref{eq:DTLN} by an ${\cal O}(1)$ rescaling, which in particular does not exclude that the actual effect on the waveform will be zero. However, the logarithmic running, as shown in our approach and reproducing the prediction obtained  with a scattering amplitude computation~\cite{Saketh:2023bul}, is expected to enter the gravitational waveform.  Regardless of the absence of a proper matching procedure, in the next sections we will discuss the implication of a potentially nonvanishing TLN at $\mathcal{O}(m^{2}\omega^{2})$ on the GW waveform.

\section{Effect of the dynamical TLNs in the waveform}
\label{sec:waveform}
\noindent
In this Section, we derive the contribution of dynamical TLNs to the gravitational waveform emitted by a compact binary system.
To this end, we extend the formalism developed in Refs.~\cite{Vines:2010ca,Vines:2011ud} (see also~\cite{Abdelsalhin:2018reg}) to account for dynamical TLNs, focusing on the case of non-rotating BHs.
We restrict our analysis to the mass quadrupole moment of the compact objects, which provides the leading contribution to the gravitational waveform~\cite{Vines:2011ud}, and neglect both higher-order multipole moments and current-type multipoles.

When a compact body of mass $m$ is embedded in a test (quadrupolar, electric) tidal field $E_{\mu\rho}=C_{\mu\nu\rho\sigma}u^\nu u^\sigma$ (where $u^\mu$ is the four-velocity of the body and $C_{\mu\nu\rho\sigma}$ is the Weyl tensor), it acquires a quadrupole moment. If the tidal field is slowly varying, it is possible to write the induced quadrupole moment in terms of an expansion in the tidal field and its time $(\tau)$ derivatives to linear order as~\cite{Saketh:2023bul} 
\begin{align}
&Q_E^{\mu \nu}= - m^5 \sum_{n = 0}^\infty (-1)^n m^n \lambda_{(n)} \frac{d^n E^{\mu \nu}}{d \tau^n} \nonumber \\
& = - m^5 \lambda_{(0)} E^{\mu \nu} +  m^6 \lambda_{(1)} \dot E^{\mu \nu} 
- m^7 \lambda_{(2)} \ddot E^{\mu \nu} + \mathcal{O}(\dddot E)\,.\label{eq:genQE}
\end{align}
For simplicity, we have left implicit the index $\ell=2$ in the TLNs, since we only consider quadrupolar deformations and, following the conventions of~\cite{Saketh:2023bul}, the TLNs are assumed to be dimensionless.
The first term in this expansion is what appears in the standard  adiabatic relations between tidal field and quadrupole; the corresponding coefficient, $\lambda_{(0)}$, is related to the (quadrupolar, electric) static TLN. In particular, by performing an effective field theory matching, it is possible to derive a direct relation between the Wilson coefficient $\lambda_{(0)}$ and the TLN computed within the framework of General Relativity, $\lambda_{(0)} \propto \,_{-2}k^{(0)}_{2}$ (see for example Ref.~\cite{Hui:2020xxx} for an extensive analysis).
The second term accounts for tidal dissipation, which---in the case of BHs---is associated with tidal heating~\cite{Hartle:1973zz,Chia:2020yla,Saketh:2023bul}. The third term captures the leading-order dynamical tidal deformation of the body; its coefficient, $\lambda_{(2)}$, is the (electric, quadrupolar)  dynamical TLN. As in the static case, establishing the precise value of this coefficient requires a dedicated matching between the effective field theory and the full general relativistic calculation, as discussed in Sec.~\ref{sec:GR} for nonrotating BHs.
Finally, in Eq.~\eqref{eq:genQE}, we will neglect time derivatives of order higher than second; 
the index in parentheses indicates the order in the time-derivative expansion.

In the absence of a fully developed matching procedure within the effective field theory framework, we shall adopt the prediction derived in Sec.~\ref{sec:GR} (see Eq.~\eqref{eq:DTLN}), and assume that the coefficients $\lambda_{(2)}$ depend on the radial distance $r$, according to the following ansatz (see also Ref.~\cite{Mandal:2023hqa}, which discusses the emergence of logarithmic running in the dynamical TLNs derived via a renormalization-based approach to scattering amplitude computations)
\begin{align}
\lambda_{(2)} (r) = {\bar\lambda}_{(2)} + \beta_{(2)} \ln \left( \frac{r}{\bar{r}} \right)\,,
\label{eq:ansatzTLN}
\end{align}
where ${\bar\lambda}_{(2)}$ and $\beta_{(2)}$ are the two coefficients that should be matched to the prediction of General Relativity of Eq.~\eqref{eq:DTLN}, whereas $\bar{r}$ is some characteristic length scale, ultimately fixed in terms of the BH properties, such as its mass,  through the matching procedure with the worldline effective theory.

The motion of a non-rotating body immersed into an external gravitational field can thus be described in terms of a worldline action of the form~\cite{Goldberger:2005cd, Porto:2016zng} 
\begin{align}
S &= S_\text{\tiny PP} + \frac{1}{4} \int {\rm d} \tau \, Q_E^{\mu \nu} E_{\mu \nu} \nonumber \\
&= - m \int {\rm d} \tau + \frac{1}{4} \int {\rm d} \tau \,  Q_E^{\mu \nu} E_{\mu \nu}\,,\label{eq:WLAction}
\end{align}
where $S_\text{\tiny PP}$ is the point-particle action, expressed in terms of the proper worldline time $\tau$, whereas $Q_E^{\mu\nu}$ is the quadrupole moment, given by Eq.~\eqref{eq:genQE}, which is coupled to the external source with amplitude $E_{\mu \nu}$. In order to keep track of conservative and dissipative interactions, associated to parity even and odd terms in the expansion~\eqref{eq:genQE}, it is possible to derive an orbital energy-balance law of the form~\cite{Vines:2010ca,Vines:2011ud, Hinderer:2009ca, Abdelsalhin:2018reg}
\begin{equation}
\frac{d\mathcal{E}}{dt}=- F\,,
\label{eq:eblaw}
\end{equation}
from which one can easily derive the phase of the gravitational waveform.
In Eq.~\eqref{eq:eblaw}, the binding energy $\mathcal{E}$ is obtained from the conservative part of the action, while the dissipative part contributes to the flux $F$. In the following, we shall only compute the contribution of the static and dynamical TLNs, neglecting the one associated to dissipative tidal heating, since the contribution of the latter to the waveform is well known, see~\cite{Hartle:1973zz,Hughes:2001jr,Maselli:2017cmm,HegadeKR:2024agt,Chia:2024bwc} and Refs therein. 

Let us now consider a binary system of two nonrotating BHs with masses $m_1$, $m_2$. Following~\cite{Vines:2010ca}, we define the mass ratios $\eta_A=m_A/M$ ($A=1,2$), where $M=m_1+m_2$ is the total mass, the symmetric mass ratio $\eta=\eta_1\eta_2$, and the reduced mass $\mu=\eta M$. We also introduce a harmonic, conformally Cartesian coordinate system $(t,x^i)$ ($i=1,\dots,3$), which we call the ``global frame''. Latin indices $i,j,\dots$ run over the three-dimensional spatial coordinates and are contracted with the Euclidean flat metric $\delta^{ij}$ (thus setting no distinction between upper and lower indices). In this frame, the worldlines of the two bodies are given by the functions $x^i=z^i_A(t)$ ($A=1,2$). We shall describe the motion of the binary, in the center-of-mass frame, in terms of the orbital separation $z^i=z^i_2-z^i_1$, the velocity $v^i={\dot z}^i$ and the acceleration $a^i={\ddot z}^i$. We also introduce polar coordinates in the plane of orbital motion, $(r,\varphi)$, where $r=\sqrt{z^iz^i}$ is the radial separation, the angular velocity $\omega=\dot\varphi$, and the unit vector $n^i=z^i/r$. Thus, $\dot r=n^iv^i$ and $v^iv^i={\dot r}^2+r^2\omega^2$.

Following Refs.~\cite{Vines:2010ca,Vines:2011ud,Abdelsalhin:2018reg}, we will describe the system using the so-called ``$m_1-m_2-Q_E$'' truncation: the first body of the system is characterized by its monopole mass moment only, while the second is characterized by its monopole mass moment and its quadrupole mass moment; all the other moments identically vanish. Once the gravitational waveform is  computed, the contribution of the tidally induced quadrupole moment of body~$1$ will be easily obtained by exchanging the indices $1,2$ of the two bodies. For simplicity, we shall neglect the higher-order PN corrections to the point-particle action, but include the static tidal term (proportional to $\lambda_{(0)}$) and the dynamical tidal term (proportional to $\lambda_{(2)}$) to the corresponding leading PN orders.

The motion of the bodies is then described by the effective Lagrangian
\begin{align}
L &= \frac{1}{2} \eta M \dot{r}^2 + \frac{1}{2} \eta M r^2 \dot{\varphi}^2 + \frac{\eta M^2}{r} \frac{1}{4} m_2^5 \nonumber \\
&\times \left[ \lambda_{(0)} E^{i j} E^{i j} (r) - m_2^2 \lambda_{(2)} (r) \dot{E}^{i j} \dot{E}^{i j} (r)  \right]\,,\label{eq:effL}
\end{align}
where~\cite{Vines:2011ud}
\begin{align}
E^{i j} &= 
- \frac{3m_1}{r^3} \left( n^i n^j - \frac{1}{3}\delta^{ij} \right)\,, \nonumber \\ 
\dot{E}^{i j} & 
 = \frac{m_1}{r^4} \left[ \dot r \left(15 n^i n^j - 3\delta^{ij} \right) -  3\left( v^i  n^j+ v^j n^i \right) \right]\,.
\end{align}
Using the identities
\begin{align}
E^{i j} E^{i j} & =\frac{2}{3}\left(\frac{3m_1}{r^3} \right)^2  \,, \nonumber \\
\dot{E}^{i j} \dot{E}^{i j} &= 
\left(54 \dot r^2 + 18 r^2 \omega^2 \right)\left(\frac{m_1}{r^4} \right)^2 \,,
\end{align}
which are valid for circular orbits (for which $\dot r=\ddot r=0$), the Euler-Lagrange equations give $\dot\omega=0$ and, at leading PN order and linearizing in the TLNs (including the contribution of the tidal interaction of body $2$ on body $1$ and labeling $\lambda_{(i)A}$ the $i$th TLNs of the Ath body):
\begin{align}
r(\omega) &=
\frac{M^{1/3}}{\omega^{2/3}} \left[ 1 + 3 \frac{(m_1 m_2^4) \omega^{10/3}}{M^{5/3}} \left(\lambda_{(0)2}  
\right.\right.\nonumber\\
&\left.\left.- 3m_2^2 \lambda_{(2)2} \omega^2
+ \frac{1}{2} m_2^2 M^{1/3} \omega^{4/3} \frac{d \lambda_{(2)2}}{dr} \right) \right.\nonumber\\
&\left.+  (1 \leftrightarrow 2)\right]\,.\label{eq:rw}
\end{align}
The binding energy then reads
\begin{align}
\mathcal{E} & = \dot{r} \frac{d L}{d \dot{r}} + \omega \frac{d L}{d \dot \varphi} - L \overset{\dot r = 0}{=} \omega \frac{d L}{d \dot \varphi} - L \nonumber \\
& = \frac{1}{2} \eta M r^2 \omega^2 - \frac{\eta M^2}{r}
- \left[ \frac{3}{2} (m_1^2 m_2^5)\lambda_{(0)2} \frac{1}{r^6} \right.
\nonumber \\
&\left. + \frac{9}{2} (m_1^2 m_2^7)\lambda_{(2)2} \frac{\omega^2}{r^6}  + (1 \leftrightarrow 2) \right]\,.
\end{align}
By replacing Eq.~\eqref{eq:rw} in the binding energy, we find
\begin{align}
    \mathcal{E} &=-\frac{1}{2} \eta M^{5/3}  \omega^{2/3}  \left\{1 - 9\frac{(m_1 m_2) \omega^{10/3}}{M^{5/3}} \right. \nonumber \\
    & \left.  \times 
   \left[ (m_1^3 \lambda_{(0)1} + m_2^3 \lambda_{(0)2}) 
    - 5 (m_1^5 \lambda_{(2)1} + m_2^5 \lambda_{(2)2})  \omega^2  \right. \right.  \nonumber \\ 
    & \left. \left. +  \frac{2}{3} \left(m_1^5 \frac{d \lambda_{(2)1}}{dr}  +m_2^5 \frac{d \lambda_{(2)2}}{dr} 
\right)  M^{1/3}\omega^{4/3}  \right] \right\}\,.
\end{align}
Let us stress that the dynamical TLNs can be a generic function of $r=r(\omega)$, following on the definition adopted in Eq.~\eqref{eq:ansatzTLN}.

By replacing the ansatz~\eqref{eq:ansatzTLN}, and linearizing to leading order in the TLNs [{\it i.e.}, using the Newtonian result when replacing $r=r(\omega)$], one gets
\begin{align}
\frac{d \lambda_{(2)A}}{dr} \Bigg|_{r = r(\omega)} = \beta_{(2)A} \frac{1}{r}\Bigg|_{r = r(\omega)} = 
\beta_{(2)A} \frac{\omega^{2/3}}{M^{1/3}}\,.
\end{align}
Thus, the binding energy reads
\begin{widetext}
\begin{align}
\mathcal{E}(\omega)& = -\frac{1}{2} \eta M^{5/3}  \omega^{2/3}  \left\{1 - 9\frac{(m_1 m_2) \omega^{10/3}}{M^{5/3}} \left[(m_1^3 \lambda_{(0)1} + m_2^3 \lambda_{(0)2}) - 5 (m_1^5 \bar \lambda_{(2)1} + m_2^5 \bar \lambda_{(2)2})  \omega^2 \right. \right. \nonumber \\
& \left. \left.+  \frac{2}{3} (m_1^5 \beta_{(2)1} +m_2^5 \beta_{(2)2}) \omega^{2}   - 5 \left(m_1^5 \beta_{(2)1} \ln \left(\frac{M^{1/3}}{\omega^{2/3}\bar{r}_1} \right)  + m_2^5 \beta_{(2)2} \ln \left(\frac{M^{1/3}}{\omega^{2/3}\bar{r}_2} \right) \right)  \omega^2  
 \right]\right\}\,.\label{eq:ew}
\end{align}
\end{widetext}
Let us now compute the GW flux emitted by the binary. At leading PN order, it is given by~\cite{Vines:2011ud}
\begin{align}
F = \frac{1}{5} \frac{d^3 M^{ij}}{d t^3} \frac{d^3 M^{ij}}{d t^3}\,,
\end{align}
where
\begin{equation}
     M^{ij} = Q_1^{ij}+Q_2^{ij} + \mu r^2 \left(n^i n^j - \frac{1}{3}\delta^{ij}\right)
\end{equation}
is the total quadrupole moment of the system. Its third time derivatives are (for circular orbits):
\begin{align}
\frac{d^3 M^{ij}}{d t^3} &= \frac{d^3 Q_1^{ij}}{d t^3}  + \frac{d^3 Q_2^{ij}}{d t^3}  \nonumber\\
&+\mu r  \left[ \dddot x^i n^j + 2w^i \dot n^j  + v^i \ddot n_j + (i \leftrightarrow j)  \right]\,,
\end{align}
where (and similarly for body 2)
\begin{align}
\frac{d^3 Q_1^{ij}}{d t^3} & = 3 (m_1^5 m_2) \lambda_{(0)1} \frac{1}{r^4} \left[ \dddot x^i  n^j +2\ddot x^i \dot n^j  + \dot x^i \ddot n^j \right]\nonumber \\
&+3 (m_1^7 m_2) \lambda_{(2)1} \frac{1}{r^4}  \left[ (\partial_t^5 x^i)  n^j + 4\ddddot x^i \dot n^j \right.\nonumber\\
&\left.+ 6\dddot x^i \ddot n^j + 4\ddot x^i \dddot n^j  + \dot x^i \ddddot  n^j\right] + (i \leftrightarrow j )\,.
\end{align}
Using the identities (again valid only for circular orbits)
\begin{align}
&x_i n_i  = r\,, \nonumber \\
&\dot x_i n_i =  x_i \dot n_i = 0\,, \nonumber \\
& \ddot x_i n_i = -\dot x_i \dot n_i = - \dot x_i \dot x_i/r = - r \omega^2\,, \nonumber \\
&  \dddot x_i n_i = \ddot x_i \dot n_i = \dot x_i \ddot n_i = 0 \,, \nonumber \\
&\ddddot x_i n_i = -\dddot x_i \dot n_i = \ddot x_i \ddot n_i = r\omega^4\,, \nonumber \\
&  (\partial_t^5 x_i) n_i= \ddddot x_i \dot n_i = \dddot x_i \ddot n_i = 0\,, 
\end{align}
one finds
\begin{widetext}
\begin{align}
F (\omega) & = \frac{32}{5} \mu^2 M^{4/3} \omega^{10/3} \Bigg\{1 + \left[m_1^4 \left(12 \frac{m_2}{M}+6\right) \lambda_{(0)1} + m_2^4 \left(12 \frac{m_1}{M}+6\right) \lambda_{(0)2} \right]\frac{\omega^{10/3}}{M^{2/3}}  \nonumber \\
&  + 12 \left[m_1^6 \left(\frac{m_2}{M}\frac{1}{2} \beta_{(2)1}- \left(3\frac{m_2}{M} +2 \right)\bar{\lambda}_{(2)1}  \right) + m_2^6 \left( \frac{m_1}{M}\frac{1}{2} \beta_{(2)2} - \left(3\frac{m_1}{M} +2 \right)\bar{\lambda}_{(2)2} \right)\right] \frac{\omega^{16/3}}{M^{2/3}}  \nonumber \\
& - 12 \left[m_1^6 \left(3\frac{m_2}{M} + 2 \right) \beta_{(2)1} \ln \left(\frac{M^{1/3}}{\omega^{2/3}\bar{r}_1} \right) + m_2^6 \left(3\frac{m_1}{M} + 2 \right) \beta_{(2)2} \ln \left(\frac{M^{1/3}}{\omega^{2/3}\bar{r}_2} \right) \right] \frac{\omega^{16/3}}{M^{2/3}} \Bigg\}\,.
\end{align}
\end{widetext}
In the stationary phase approximation, the phase contribution can then be derived from~\cite{Vines:2011ud}
\begin{equation}
\frac{d^2 \psi}{d \omega^2} = - \frac{2}{F} \frac{d \mathcal{E}}{d \omega}\,.
\end{equation}
By combining the expression for $F$ and $d\mathcal{E}/d\omega$, and performing the integration in the frequency domain, one finally gets the phase of the gravitational waveform, which in terms of the dimensionless frequency $x = (M \omega)^{2/3}$ reads:
\begin{widetext}
\begin{align}
\label{psi-gen}
\psi(x) &= \frac{3}{128 \, \eta \, x^{5/2}} \left\{1 -\frac{24}{M^4}
 \left[m_1^3(m_1 +11 \mu)\lambda_{(0)1} + m_2^3(m_2 + 11 \mu)\lambda_{(0)2} \right] x^{5} \right. \nonumber \\
& + \frac{15}{11 M^6}  \left[m_1^5(8m_1 +147 \mu)\bar \lambda_{(2)1}+m_2^5(8m_2 +147 \mu)\bar \lambda_{(2)\,2}- 35 m_1^5\mu \beta _{(2)\,1} - 35 m_2^5  \mu \beta _{(2)2}   \right] x^{8} 
\nonumber\\
& \left. + \frac{15}{11M^6} \left[  m_1^5(8m_1 + 147 \mu) \beta_{(2)1} \ln \left(\frac{M}{\bar{r}_1 x}\right)  + m_2^5(8m_2 + 147 \mu)\beta _{(2)2} \ln \left(\frac{M}{\bar{r}_2 x}\right)\right] x^{8}  
\right\}\,.
\end{align}
\end{widetext}
By collecting the  contributions entering at different PN terms, we can rewrite the phase in the more compact form as\footnote{Let us emphasize that there is an intrinsic ambiguity associated with the logarithmic term in the tidal PN phase. In particular, the effective renormalization scales \( \bar{r}_A \) can, in principle, be absorbed into the 8PN tidal coefficient \( \Lambda_{(2)} \), effectively leaving a residual \( \sim \ln x \) contribution in the \( B_{(2)A} \) terms. This separation appears more natural from a PN standpoint. However, due to the absence of a proper matching procedure with the underlying effective theory---which would determine \( \bar{r}_A \) in terms of the BH radius---we choose to remain agnostic about this ambiguity. Accordingly, we retain the logarithmic term as part of the next-to-leading-order correction to the dynamical TLN contribution in the waveform.}
\begin{align}
\label{phase}
\psi (x) &\equiv \frac{3}{128 \, \eta \, x^{5/2}} \left[1 -\frac{39}{2}\Lambda_{(0)} x^5  + \frac{15}{11}  \Lambda_{(2)}x^{8} \right. \nonumber \\
& \left.  + \frac{15}{11} B_{(2)1} \, x^{8}  \ln \left(\frac{M}{\bar{r}_1 x} \right) +\frac{15}{11} B_{(2)2} \, x^{8}  \ln \left(\frac{M}{\bar{r}_2 x} \right) 
\right]\,,
\end{align}
where the coefficients $\Lambda_{(0)}, \Lambda_{(2)}, B_{(2)A}$ can be read off from Eq.~\eqref{psi-gen}.
The above equation is the main result of this Section: it provides the GW phase corrections from the dynamical TLNs
of a BH, including putative logarithmic terms associated with the running of the TLNs. Since the static TLNs of a BH are strictly zero $\Lambda_{(0)}=0$, and the corrections proportional to $\Lambda_{(2)}$ and $B_{(2)A}$ are also the first nonvanishing conservative tidal contributions in a BH binary signal.\footnote{In the above derivation we are ignoring the dissipative tidal correction (tidal heating) entering at 2.5PN$\times \ln x$ or 4PN$\times\ln x$ for spinning or nonspinning objects, respectively~\cite{Hughes:2001jr}.}
As expected by the fact that the dynamical TLNs are quadratic in the frequency, their phase contribution is suppressed by an extra $\omega^2M^2\sim M^3/r^3$ factor compared to the usual (static) tidal contribution. Since the latter enters at 5PN order in the GW phase~\cite{Flanagan:2007ix}, the dynamical tidal terms $\lambda_{(2)A}$ enter at 8PN~\cite{Pitre:2023xsr,Pitre:2025qdf}, while the putative logarithmic corrections are suppressed by an extra $\ln(x)$ term.

\section{Detectability of the dynamical TLNs of a BH}
\label{sec:Fisher}
\noindent
To forecast the measurement precision of dynamical tidal parameters in the waveform, we adopt the Fisher information matrix formalism, a well-established tool for analyzing high signal-to-noise ratio (SNR) signals~\cite{Vallisneri:2007ev}. 
We anticipate that, even within some optimistic assumption, the prospects to detect dynamical tidal effects in BHs are negative, even for the loudest signals expected with future interferometers.

We model the GW signal using the \textsc{IMRPhenomD} waveform family~\cite{Husa:2015iqa,Khan:2015jqa}, which captures the full inspiral-merger-ringdown evolution of binary coalescences. We extend this model to include dynamical tidal effects in the GW phase via Eq.~\eqref{phase}, restricting ourselves to the leading 8PN contribution $\Lambda_{(2)}$. Since we focus on binary BH systems, we impose from the outset the vanishing of the static TLNs by setting $\Lambda_{(0)}  = 0$. We also neglect tidal dissipation effects, in order to isolate and highlight the potential detectability of the 8PN dynamical tidal corrections. A more comprehensive analysis including dissipation is deferred to future work.

In the frequency domain, the waveform template is given by~\cite{Sathyaprakash:1991mt, Damour:2000gg}:
\begin{equation}\label{eq:htilde}
\tilde h (f;\bm{\theta}) = C_\Omega \, \mathcal{A}(f;\bm{\theta}) \, e^{i [\psi_\text{\tiny PP}(f;\bm{\theta}) + \psi_{\rm tidal}(f;\bm{\theta})]}\,,
\end{equation}
where $\bm{\theta}$ denotes the set of waveform parameters, including both intrinsic and extrinsic properties of the binary. The amplitude $\mathcal{A}(f;\bm{\theta})$ and the point-particle phase $\psi_\text{\tiny PP}$ are provided by the spin-aligned \textsc{IMRPhenomD} model. 

The leading-order amplitude takes the form~\cite{Berti:2004bd}:
\begin{equation}
\mathcal{A} (f;\bm{\theta}) = \sqrt{\frac{5}{24}} \frac{\mathcal{M}^{5/6}}{\pi^{2/3} \, d_L \, f^{7/6}}\,,
\end{equation}
where $\mathcal{M}$ is the chirp mass and $d_L$ is the luminosity distance. The geometric factor $C_\Omega$ captures the detector's response and depends on the binary’s inclination angle as well as the antenna pattern functions, which in turn vary with the source's sky location, polarization angle, and frequency. For simplicity, we assume optimally oriented binaries, thereby neglecting these dependencies in our analysis.

To estimate how well parameters can be measured, we compute the Fisher matrix~\cite{Cutler:1994ys, Poisson:1995ef, Vallisneri:2007ev}:
\begin{equation}
\Gamma_{ij} = \left\langle \frac{\partial h}{\partial \theta_i} \bigg\vert \frac{\partial h}{\partial \theta_j} \right\rangle \bigg|_{\bm{\theta} = \bm{\hat{\theta}}}\,,
\end{equation}
where $\bm{\hat{\theta}}$ are the true parameter values. The square root of the diagonal elements of the inverse matrix, $\Sigma_{ij} = (\Gamma^{-1})_{ij}$, gives the $1\sigma$ uncertainties $\sigma_i$ on the parameters. The inner product between two waveforms $h_1$ and $h_2$ is defined as
\begin{equation}
\langle h_1 | h_2 \rangle = 4 \, \Re \int_{f_{\min}}^{f_{\max}} \frac{\tilde h_1^*(f) \tilde h_2(f)}{S_n(f)} \, df,
\end{equation}
with $S_n(f)$ the detector’s noise power spectral density. 
We define the SNR as $\rho = \sqrt{\langle h | h \rangle}$. The frequency range of the integral depends on the detector. For the ET, we use $f_{\min} = 2$ Hz and $f_{\max} = 4096$ Hz; for LISA, we set $f_{\min} = 10^{-5}$ Hz and $f_{\max} = 0.3$ Hz, assuming a 4-year mission duration. Detector noise curves follow Refs.~\cite{Abac:2025saz,Branchesi:2023mws,Babak:2021mhe}.

Our full waveform thus depends on seven parameters:
$\bm{\theta} = \{{\cal M}, \eta, \chi_s, \chi_a, t_c, \phi_c, \Lambda_{(2)}\}$, fixing $(t_c, \phi_c) = (0, 0)$ for convenience. We consider equal-mass, nonspinning ($\chi_s= \chi_a =0$) binary BH systems with component masses $m_1 = m_2 = m_{\text{\tiny\text{\tiny BH}}}$ at a luminosity distance of $d_L = 100 \, {\rm Mpc}$ and $d_L = 1 \, {\rm Gpc}$ for ET and LISA, respectively. 
These are very optimistic choices to maximize the event SNR.
To account for the uncertainties in the exact value of the dynamical TLNs, we span a few values for the dynamical tidal coefficients $\bar \lambda_{(2)1} = \bar \lambda_{(2)2} \equiv \bar \lambda_{(2)}$ and $\beta_{(2)1} = \beta_{(2)2} \equiv \beta_{(2)}$, including also those obtained in Sec.~\ref{sec:GR} in the context of General Relativity which, even though lacking a proper matching procedure, we expect to be of the same order of magnitude.

\begin{figure*}[tbp] 
    \includegraphics[width=0.99\textwidth]{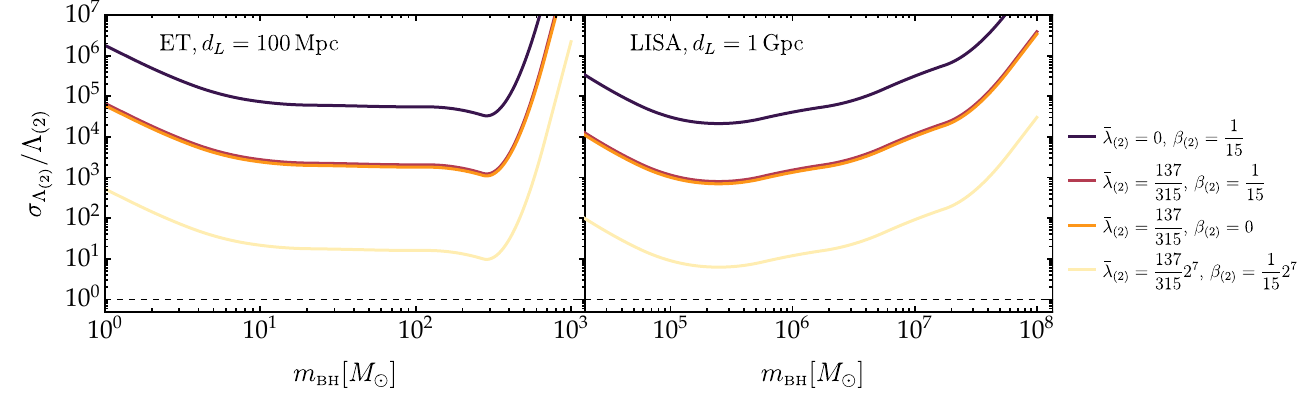}
    \caption{Relative error on the dynamical TLN parameter $\Lambda_{(2)}$, for an equal-mass, nonspinning binary BH system with individual masses $m_\BH$, observed by ET (left panel) and LISA (right panel) optimistically assuming a luminosity distance of $d_L = 100\,{\rm Mpc}$ and $d_L = 1\,{\rm Gpc}$, respectively. The colored lines identify different choices of values of the dynamical TLN coefficients $\bar \lambda_{(2)}, \beta_{(2)}$. A measurement with $100\%$ accuracy would correspond to $\sigma_{\Lambda_{(2)}}/\Lambda_{(2)}=1$ (dashed horizontal line).
    }
\label{fig:Fisher}
\end{figure*}

The results of the analysis  are shown in Fig.~\ref{fig:Fisher}, where we plot the statistical uncertainty on the dynamical TLN parameter $\Lambda_{(2)}$ in terms of the BH mass.

As evident from the plots, even within our simplified waveform and distance assumptions, the statistical errors on $\Lambda_{(2)}$ are very large. A measurement with $100\%$ accuracy would correspond to $\sigma_{\Lambda_{(2)}}/\Lambda_{(2)}=1$, which is at least three orders of magnitude below the curves shown in Fig.~\ref{fig:Fisher} for  $\lambda_{(2)}=\,_{-2}k^{(2)}_{2}$. This also shows that, should $\lambda_{(2)}$ be a factor ${\cal O}(100)$ larger than $\,_{-2}k^{(2)}_{2}$, our conclusions are expected to remain valid.\footnote{As we shall mention in the concluding Section, calculations of dynamical TLNs for neutron stars suggest that $\lambda_{(2)}\sim(R/m)^8 \,_{-2}k^{(2)}_{2}$~\cite{Pitre:2023xsr}, where $R$ is the characteristic radius of the object (in fact, order-of-magnitude estimates based on the effective theory suggest that $\lambda_{(2)}=(R/m)^7 \,_{-2}k^{(2)}_{2}$). For a BH, one would expect $R_\text{\tiny BH}\approx 2m$ and the dynamical tidal coefficient in the waveform can be amplified by a ${\cal O}(100)$ factor compared to the dynamical TLN $\,_{-2}k^{(2)}_{2}$ computed within BH perturbation theory. As shown in Fig.~\ref{fig:Fisher}, our conclusions about the non-detectability would remain valid also in this case.}
Within the Fisher approximation, the errors scale inversely with the SNR and hence linearly with the distance, so the prospect is clearly even more pessimistic for more distant events.
Furthermore, we are neglecting spin precession and eccentricity, the inclusion of which would increase the number of waveform parameters, presumably increasing the errors and potentially introducing parameter correlations. Finally, we considered only a collective dynamical tidal coefficient $\Lambda_{(2)}$ in the waveform, which depends on the two individual dynamical TLNs of the binary components. Measuring the latter independently requires including subleading PN point-particle contributions, which are still unknown from  5PN order up to 8PN order, thus making the measurement even more challenging. 

\section{Conclusion and discussion}
\label{conclusions}
\noindent
In this work, we have investigated the dynamical tidal response of Schwarzschild BHs both within the framework of BH perturbation theory and PN theory, and perturbatively in the frequency of the tidal field, focusing on the conservative and dissipative contributions up to quadratic order in the frequency. By employing the Teukolsky formalism in advanced null coordinates, we derived a particularly clean perturbative scheme that can be recast in a simple iterative form. 
As a possible extension, it would be interesting to check if one can solve the equations at any perturbative order and eventually resum the small-frequency series.
Furthermore, by defining the tidal response function in terms of the full solution to the perturbation equations at quadratic order, we obtain a response that is independent of the arbitrary constants of the homogeneous solutions at lower order, unlike previous approaches.  

Our analytical computations agree with past results for the leading-order dissipative response appearing at linear order in frequency. Importantly, within perturbation theory, we have found indication that the second-order dynamical TLNs are nonvanishing and possess a logarithmic running dependence, compatible with scattering-amplitude results. Once properly matched to the worldline action coefficients, our method allows for an unambiguous determination of the tidal response in a particularly clean way.

Although we did not compute the exact matching explicitly---postponing such computation to future work---we could estimate the effect of the dynamical TLNs within ${\cal O}(1)$ ambiguities. Based on this consideration, we further explored the observational impact of dynamical TLNs by computing the corresponding corrections to the GW phase, which appear at the 8PN order. These corrections, although theoretically well motivated, are highly suppressed, making them effectively unmeasurable even with optimistic assumptions and for third-generation detectors like ET or space-based missions such as LISA.

One should also consider that the tidal response of a BH, even if possibly nonzero, is not amplified by high powers of the inverse compactness $R/m$ as in the neutron star case (here $R\approx 10m$ is the radius of the star). The static TLNs, which are zero for BHs, are amplified by a factor $(R/m)^5$~\cite{Hinderer:2007mb}, whereas the dynamical tidal terms are amplified by a factor $(R/m)^8$~\cite{Pitre:2023xsr} (see also \cite{Mandal:2023hqa}), although also in that case a proper matching to the waveform parameters is lacking and might change this scaling\footnote{Indeed, order-of-magnitude estimates suggest a factor $(R/m)^7$.}. Thus, although suppressed by a high PN order, for neutron stars the dynamical TLNs might be still relevant to faithfully model the GW signal and reduce systematics~\cite{Mandal:2023lgy}\footnote{We stress that, for neutron stars, the low-frequency expansion of the dynamical tidal response can become inaccurate in the late inspiral, where internal fluid modes are excited and lead to strong resonant behavior. A more complete treatment of these effects is thus required in order to assess the strength of dynamical tides during the late inspiral of neutron stars (see, e.g., Ref.~\cite{Steinhoff:2016rfi} for work along this direction).
}.
We plan to study this interesting problem in future work.

Although we cannot exclude that, also for BHs, the tidal waveform parameters are significantly larger than the dynamical TLNs computed here within BH perturbation theory, the absence of a hierarchy of scales ({\it i.e.}, $R_\text{\tiny BH}\sim m$, unlike the neutron-star case) suggests that this is not the case.
Hence, for BH binaries, the 8PN tidal corrections to the GW phase are not parametrically larger than the subleading PN point-particle contribution entering at the same PN order. Given that those terms are currently unknown and---in the light of our results---hardly measurable, it seems that the dynamical TLNs for BHs will not have a phenomenological relevance in any foreseeable future. Nonetheless, assessing whether they really affect the waveform of a BH binary remains an interesting theoretical question. This can be unambiguously addressed by bridging our BH perturbation theory and PN results with a proper matching. We leave this prospect to future work.

\section*{Acknowledgments}
\noindent
We acknowledge interesting discussions with Vitor Cardoso, Takuya Katagiri and Kent Yagi. We also thank Miguel Correia, Giulia Isabella and Julio Parra Martinez for useful correspondence.
Research of S.C. is supported by the MATRICS and the Core research grants from SERB, Government of India (Reg. Nos. MTR/2023/000049 and CRG/2023/000934).
V.DL. is supported by funds provided by the Center for Particle Cosmology at the University of Pennsylvania.
P.P. is partially supported by the MUR FIS2 Advanced Grant ET-NOW (CUP:~B53C25001080001), by the PRIN Grant 2020KR4KN2 ``String Theory as a bridge between Gauge Theories and Quantum Gravity'', by the FARE programme (GW-NEXT, CUP:~B84I20000100001), and by the INFN TEONGRAV initiative. 
We acknowledge financial support from the EU Horizon 2020 Research and Innovation Programme under the Marie Sklodowska-Curie Grant Agreement No. 101007855.

\appendix

\section{Dynamical tidal response of a Schwarzschild BH for generic spin-$s$ perturbations}\label{app:genericspin}
\noindent
In this Appendix, we compute the tidal response of Schwarzschild BHs to generic spin-$s$ perturbations. Following the definitions introduced in Eq.~\eqref{PBA}, the radial perturbation equation in this case depends on the following coefficients: $P_{-} = B_{-} = 0$, $P_{+} = -2m\omega$, $B_{+} = 4m\omega$, and $A_{+} = A_{-} = \ell(\ell+1) - s(s+1)$. With these expressions, the radial equation for generic spin-$s$ perturbations takes the form:
\begin{align}\label{gen_rad}
&\dfrac{d^{2}\,_{s}R_\ell}{dz^{2}}+\left[\frac{(1+s)-4im\omega}{z}+\frac{1+s}{1+z}-4im\omega\right]\dfrac{d\,_{s}R_\ell}{dz}
\nonumber
\\
&+\Big[-\frac{4im\omega s}{z^{2}}-\frac{E_{\ell m_{z}}+4im\omega}{z}+\frac{E_{\ell m_{z}}}{(1+z)}\Big]\,_{s}R_\ell=0\,,
\end{align}
where we have removed the $m_z$ from the subscript of the radial perturbation because of spherical symmetry. The above differential equation, when multiplied by $z(1+z)$, can be expressed in the following form
\begin{align}
&z(1+z)\dfrac{d^{2}\,_{s}R_\ell}{dz^{2}}+\left[(1+s)(1+2z)-4im\omega(1+z)^{2}\right]\dfrac{d\,_{s}R_\ell}{dz}
\nonumber
\\
&-\Big[4im\omega\left(\frac{s}{z}+1+z+s\right)+E_{\ell m_{z}}\Big]\,_{s}R_\ell=0\,.
\end{align}
Expanding the radial perturbation as a power series in $m\omega$, such that $\,_{s}R_\ell=\,_{s}R^{(0)}_\ell+m\omega \,_{s}R^{(1)}_\ell+m^{2}\omega^{2}\,_{s}R^{(2)}_\ell$, the above equation can be decomposed into a system of equations:
\begin{align}
\mathcal{D}^{(s)}_{0}\,_{s}R^{(0)}_\ell&=0\,,
\label{zer_gen_spin}
\\
\mathcal{D}^{(s)}_{0}\,_{s}R^{(1)}_\ell&=4i\mathcal{D}^{(s)}_{1}\,_{s}R^{(0)}_\ell\,,
\label{recgens1}
\\
\mathcal{D}^{(s)}_{0}\,_{s}R^{(2)}_\ell&=4i\mathcal{D}^{(s)}_{1}\,_{s}R^{(1)}_\ell\,,
\label{recgens2}
\end{align}
which continues iteratively as in the $s=-2$ case,
$\mathcal{D}^{(s)}_{0} \,_{s}R^{(j)}_\ell=4i\mathcal{D}^{(s)}_{1} \,_{s}R^{(j-1)}_\ell$.
The two differential operators, $\mathcal{D}^{(s)}_{0}$ and $\mathcal{D}^{(s)}_{1}$, defining the perturbative equations order by order, have the following explicit expressions: 
\begin{align}
\mathcal{D}^{(s)}_{0}&=z(1+z)\dfrac{d^{2}}{dz^{2}}+(1+s)(1+2z)\dfrac{d}{dz}-E_{\ell m_{z}}\,,
\\
\mathcal{D}^{(s)}_{1}&=(1+z)^{2}\dfrac{d}{dz}+\left(\frac{s}{z}+s+1+z\right)\,.
\end{align}
The zeroth-order equation for the radial perturbation, presented in Eq.~\eqref{zer_gen_spin}, can be solved exactly, with the following solution,
\begin{align}\label{zero_sol_gens}
\,_{s}R^{(0)}_\ell=\left\{z(1+z)\right\}^{-s/2}\left[c_{1}P_{\ell}^{|s|}(1+2z)+c_{2}Q_{\ell}^{|s|}(1+2z)\right]\,,
\end{align}
which reduces to Eq.~\eqref{sol_zero} for $s=-2$. The above associated Legendre polynomials can be expressed in terms of hypergeometric functions, which can be further manipulated to yield the following expression for the zeroth radial perturbation function
\begin{widetext}
\begin{align}\label{gens_resp_tide}
\,_{s}R^{(0)}_\ell=f^{\frac{|s|-s}{2}}\left(\frac{2m}{r}\right)^{s}\Bigg[\mathcal{A}_{s}&\underbrace{\left(\frac{r}{2m}\right)^{\ell}\,_{2}F_{1}\left(|s|-\ell,-\ell,-2\ell;\frac{2m}{r}\right)}_{\rm tidal\, part}
+\mathcal{B}_{s}\underbrace{\left(\frac{r}{2m}\right)^{-\ell-1}\,_{2}F_{1}\left(|s|+1+\ell,\ell+1,2+2\ell;\frac{2m}{r}\right)}_{\rm response}\Bigg]\,.
\end{align}
\end{widetext}
The above depicts the decomposition of the zeroth radial perturbation in terms of the tidal and response parts for generic spin perturbations. For $s=-2$, it is simple to check that the above equation reduces to Eq.~\eqref{tiderespzero}. Furthermore, the constants $\mathcal{A}_{s}$ and $\mathcal{B}_{s}$ have the following expressions in terms of the arbitrary constants $(c_{1},c_{2})$: 
\begin{align}
\mathcal{A}_{s}&=c_{1}\frac{\Gamma(1+2\ell)}{\Gamma(1+\ell)\Gamma(\ell-|s|+1)}\,,
\\
\mathcal{B}_{s}&=\frac{\Gamma(\ell+|s|+1)\Gamma(1+\ell)}{2\Gamma(2+2\ell)}\left(c_{2}+(-1)^{|s|}c_{1}\frac{\tan(\pi \ell)}{\pi} \right)\,.
\end{align}
Proceeding further, like the gravitational case, it follows that the associated Legendre polynomial $Q_{\ell}^{|s|}$, which is one of the solutions of the zeroth-order radial perturbation equation, involves a $\ln(r-2m)$ term and hence must be absent in the BH limit. Thus, regularity of perturbations at the BH horizon demands $c_{2}=0$. The zeroth-order perturbation for a Schwarzschild BH, in the asymptotic limit, reads (assuming $\ell$ to be an integer),
\begin{align}
\,_{s}R^{(0)}_\ell\sim r^{\ell-s}\,,
\end{align}
where we have used the result that $\,_{2}F_{1}(a,b,c;0)=1$ in Eq.~\eqref{gens_resp_tide}. Since $\ell\geq |s|$, it follows that the zeroth-order radial perturbation will be growing for all values of $\ell$, and hence, following the definition of the response function and the absence of any decaying mode, one is able to derive the relations 
\begin{align}\label{staticTLN}
\,_{s}k_{\ell}=0+\mathcal{O}(m\omega)\,;\qquad \,_{s}\nu_{\ell}=0+\mathcal{O}(m\omega)\,.
\end{align}
These show that, for generic spins $s$ and angular multipole $\ell$, the static tidal deformability of a Schwarzschild BH vanishes. 

At first-order in the frequency, it is difficult to obtain an analytic solution of Eq.~\eqref{recgens1} for generic $s$ and $\ell$, and it is even more difficult to obtain the second-order radial function from Eq.~\eqref{recgens2}. Thus, in order to provide explicit solutions, in the following we will need to fix the $s$ and the $\ell$ values to obtain analytical radial solutions and hence determine the dynamical TLNs. While in the main text we have already worked out the case of $s=-2$, here we depict the case of $s=0$ ({\it i.e.}, considering scalar perturbations). In this case, the zeroth-order radial perturbation, which is regular at the BH horizon,  reads (fixing $\ell = 2$)
\begin{align}
\,_{0}R^{(0)}_{2}&=c_1\left(6z^{2}+6z+1\right)\,.
\end{align}
Substituting the zeroth-order solution in Eq.~\eqref{recgens1}, we obtain the following expression for the first-order radial perturbation: 
\begin{align}
&\,_{0}R^{(1)}_{2}=\left(6z^{2}+6z+1\right)\left[i c_{3}+\frac{ic_{4}}{2}\ln \left(\frac{1+z}{z}\right)\right]
\nonumber
\\
&-\frac{3ic_{4}}{2}(1+2z)+2ic_{1}\left\{z\left[13+6z(4+z)\right] 
\right. \nonumber
\\
&\left. +14[1+6z(1+z)]\ln 2+[1+6z(1+z)]\ln(1+z)\right\}\,.
\end{align}
Similarly to the derivation in the main text, the first-order radial perturbation can be written as a linear combination of the homogeneous solutions, characterized by two arbitrary constants, \( (c_3, c_4) \), while the coefficient \( c_1 \) (appearing at zeroth-order) multiplies the particular solution. Imposing regularity at the BH horizon sets \( c_4 = 0 \), thereby simplifying the first-order solution, which in the asymptotic limit yields 
\begin{align}
&\,_{0}R^{(0)}_{2}+m\omega \,_{0}R^{(1)}_{2}\approx 12im\omega c_{1}\left(\frac{r}{2m}\right)^{3}+6\Big\{c_{1}+8i c_1 m\omega 
\nonumber
\\
&+i c_{3}m\omega +28 i c_{1}m\omega \ln 2 +2ic_{1} m\omega \ln\left(\frac{r}{2m}\right)\Big\}\left(\frac{r}{2m}\right)^{2}
\nonumber
\\
&+\cdots+\frac{im\omega c_{1}}{15}\left(\frac{2m}{r}\right)^{3}+ \mathcal{O}(r^{-4})\,.
\end{align}
The presence of a decaying part induced by the first-order solution allows one to extract a nonvanishing response function (see Eq.~\eqref{defResponse})
\begin{align}
\,_{0}\mathcal{F}_{2}=\frac{im\omega}{90}+\mathcal{O}(m^{2}\omega^{2})\,,
\end{align}
which is independent of any arbitrary constants in the solution. 
Therefore, the tidal dissipation number reads
\begin{align}
\,_{0}\nu_{2}=\frac{m\omega}{90}+\mathcal{O}(m^{3}\omega^{3})\,.
\end{align}
Finally, substituting the first-order solution in Eq.~\eqref{recgens2}, one can estimate the second-order radial perturbation. Due to its involved, and not particularly illuminating, expression, we do not present it here. However, when plugged in the total radial perturbation, one can extract the asymptotic behavior, getting
\begin{widetext}
\begin{align}
&\,_{0}R_{2} \approx - \frac{96}{7} c_1 m^2 \omega^2 \left( \frac{r}{2m} \right)^4 -\frac{4}{7} \left( \frac{r}{2m} \right)^3 \left[42 c_1 m^2 \omega^2 \ln \left( \frac{r}{2m} \right)+ 588 c_1 m^2 \omega^2 \ln 2 +167 c_1 m^2 \omega^2+21  c_3 m^2 \omega^2-21 i c_1 m \omega\right] \nonumber \\
& +\frac{2}{105} \left( \frac{r}{2m} \right)^2 \left[-630 c_1 m^2 \omega^2 \ln^2\left( \frac{r}{2m} \right)-5514 c_1 m^2 \omega^2 \ln \left( \frac{r}{2m} \right)-630  c_3 m^2 \omega^2 \ln \left( \frac{r}{2m} \right) +6090 \pi ^2 c_1 m^2 \omega^2 +315 c_1\right. \nonumber \\
& \left.-12428 c_1 m^2 \omega^2+1470  \pi ^2 c_3 m^2 \omega^2-2520  c_3 m^2 \omega^2+315 c_5 m^2 \omega^2+630 i c_1 m \omega \ln \left( \frac{r}{2m} \right)+2520 i c_1 m \omega+315 i c_3 m \omega \right. \nonumber \\
& \left. -17640 c_1 m^2 \omega^2 \ln 2 \ln \left( \frac{r}{2m} \right) -70560 c_1 m^2 \omega^2 \ln 2 +5880 \pi ^2 c_1 m^2 \omega^2  \ln 128 +8820 i c_1 m \omega \ln 2
\right] \nonumber \\
& + \dots + \frac{-363 c_1 m^2 \omega^2-70 c_3 m^2 \omega^2+70 i c_1 m \omega  -1960  c_1 m^2 \omega^2 \ln 2}{1050} \left( \frac{r}{2m} \right)^{-3} + \mathcal{O}(r^{-4})\,. \end{align}
\end{widetext}
The response function for scalar perturbations up to quadratic order in the frequency then reads 
\begin{align}
\,_{0}\mathcal{F}_{2}=\frac{im\omega}{90}+\frac{m^{2}\omega^{2}}{45}\left[\ln\left(\frac{r}{2m}\right)+\frac{197}{140} \right]\,,
\end{align}
which implies a potentially nonvanishing conservative dynamical TLNs, 
\begin{align}
\,_{0}k_{2}=\frac{m^{2}\omega^{2}}{90}\left[\ln\left(\frac{r}{2m}\right)+ \frac{197}{140}  \right]+\mathcal{O}(m^{3}\omega^{3})\,.
\end{align}
To summarize, Schwarzschild BHs have vanishing static TLNs, but potentially nonvanishing dynamical TLNs also for scalar perturbations (which, to be firmly established, requires a matching procedure with the effective theory). The same computation can be easily performed for vector ($s=\pm1$) perturbations, yielding qualitatively similar results, {\it i.e.}, $\,_{-1}k_{2}\propto\omega^2$ and $\,_{-1}\nu_{2}\propto\omega$.

\bibliography{draft}

\end{document}